\documentclass[prc,preprint,showpacs,amsmath,amssymb,nofootinbib]{revtex4-1}

\usepackage{graphicx}
\usepackage{dcolumn}
\usepackage{bm}       
\usepackage{latexsym} 
\usepackage{longtable}
\usepackage{array}
\usepackage{rotating}
\usepackage{color}
\begin{document}

\title{E2 strengths and transition radii difference of one-phonon $2^+$ states of $^{92}$Zr from electron scattering at low momentum transfer}

\author{A.~Scheikh~Obeid}
\affiliation{Institut f\"ur Kernphysik, Technische Universit\"at Darmstadt, 
                64289 Darmstadt, Germany}

\author{O.~Burda}
\affiliation{Institut f\"ur Kernphysik, Technische Universit\"at Darmstadt, 
                64289 Darmstadt, Germany}

\author{M.~Chernykh}
\affiliation{Institut f\"ur Kernphysik, Technische Universit\"at Darmstadt, 
                64289 Darmstadt, Germany}

\author{A.~Krugmann}
\affiliation{Institut f\"ur Kernphysik, Technische Universit\"at Darmstadt, 
                64289 Darmstadt, Germany}

\author{P.~von~Neumann-Cosel}
\affiliation{Institut f\"ur Kernphysik, Technische Universit\"at Darmstadt, 
                64289 Darmstadt, Germany}

\author{N.~Pietralla}
\affiliation{Institut f\"ur Kernphysik, Technische Universit\"at Darmstadt,
 64289 Darmstadt, Germany}

\author{I.~Poltoratska}
\affiliation{Institut f\"ur Kernphysik, Technische Universit\"at Darmstadt, 
                64289 Darmstadt, Germany}

\author{V.~Yu.~Ponomarev}
\affiliation{Institut f\"ur Kernphysik, Technische Universit\"at Darmstadt, 
                64289 Darmstadt, Germany}

\author{C.~Walz}
\affiliation{Institut f\"ur Kernphysik, Technische Universit\"at Darmstadt, 
                64289 Darmstadt, Germany}

\date{\today}

\begin{abstract}
\begin{description}
\item[Background] Mixed-symmetry $2^+$ states in vibrational nuclei are characterized by a sign change between dominant proton and neutron valence-shell components with respect to the fully symmetric $2^+$ state.
The sign can be measured by a decomposition of proton and neutron transition radii with a combination of inelastic electron and hadron scattering [C. Walz {\em et al.}, Phys. Rev. Lett. \textbf{106}, 062501 (2011)].
For the case of $^{92}$Zr, a difference could be experimentally established for the neutron components, while about equal proton transition radii were indicated by the data.
\item[Purpose] Determination of the ground-state (g.s.) transition strength of the mixed-symmetry $2^+_2$ state and verification of the expected vanishing of the proton transition radii difference between the one-phonon $2^+$ states in $^{92}$Zr. 
\item[Method] Differential cross sections for the excitation of one-phonon $2^+$ and $3^-$ states in $^{92}$Zr have been measured with the ($e,e'$) reaction at the S-DALINAC in a momentum transfer range $q \simeq 0.3 - 0.6$ fm$^{-1}$.
\item[Results] Transition strengths $B(E2;2^{+}_{1}\rightarrow0^{+}_{1}) = 6.18(23)$, $B(E2;2^{+}_{2}\rightarrow0^{+}_{1}) = 3.31(10)$ and
$B(E3;3^{-}_{1}\rightarrow0^{+}_{1}) = 18.4(1.1)$ Weisskopf units are determined from a comparison of the experimental cross sections to quasiparticle-phonon model (QPM) calculations.
It is shown that a model-independent plane wave Born approximation (PWBA) analysis can fix the ratio of $B(E2)$ transition strengths to the $2^+_{1,2}$ states with a precision of about 1\%. 
The method furthermore allows to extract  their proton transition radii difference. With the present data $\Delta R = -0.12(51)$ fm is obtained.  
\item[Conclusions] Electron scattering at low momentum transfers can provide information on transition radii differences of one-phonon $2^+$ states even in heavy nuclei.
Proton transition radii for the $2^+_{1,2}$ states in $^{92}$Zr are found to be identical within uncertainties.
The g.s.\ transition probability for the mixed-symmetry state can be determined with high precision limited only by the available experimental information on the $B(E2;2^{+}_{1}\rightarrow0^{+}_{1}$) value.
\end{description}
\end{abstract}

                           
\maketitle

\section{Introduction}

Collectivity, isospin symmetry and shell structure are generic features of the nuclear many-body quantum system. Collective nuclear valence-shell excitations are a key to understand how these features coexist, interplay and compete. 
In vibrational nuclei, the development of predominantly proton-neutron symmetric collective nuclear structures at low excitation energies is governed by the strong residual proton-neutron interaction. Their existence implies - due to quantum-mechanical orthogonality - the formation of collective states with, at least partial antisymmetry with respect to the contribution of proton and neutron valence-space components to their wave functions. 
Such excited states are said to have mixed symmetry~\cite{Arima01}. 
The investigation of mixed-symmetry states (MSS) is an important source of information on the effective proton-neutron interaction in the valence shell of heavy atomic nuclei~\cite{Pietralla01}. 

MSS have been defined in the framework of the proton-neutron Interacting Boson Model (IBM-2)~\cite{Iachello01}. 
In analogy to the isospin symmetry of nucleons, the symmetry of a multi-boson wave function formed by $N_{\pi}$ proton bosons and $N_{\nu}$ neutron bosons is quantified by the so-called $F$-spin. 
States with $F<F_{\rm max}=(N_{\pi}+N_{\nu})/2$ have wave functions that contain at least one pair of proton and neutron bosons antisymmetric under the exchange of proton and neutron labels. 
The signatures of $2^+$ MSS are ($i$) strong $M1$ transitions to fully symmetric states (FSS) with matrix elements of about 1$\mu^{2}_{N}$ and ($ii$) weakly collective $E2$ transitions to FSS.

The prediction of the IBM-2 with respect to a multi-phonon structure of MSS in vibrational nuclei was confirmed about ten years ago by the observation of large $M1$ transition strengths between low-energy states of $^{94}$Mo~\cite{Pietralla02,Pietralla03,Fransen01}. 
The $2^{+}$ states  were also investigated with electron scattering experiments at the superconducting electron accelerator S-DALINAC and with proton scattering at iThemba LABS~\cite{Burda01}. 
The combined analysis  supported a one-phonon structure of the $2^{+}_{1,3}$ states of $^{94}$Mo. 

In the neighboring even-even isotone $^{92}$Zr with two neutrons outside the $N = 50$ closed shell and with the $Z = 40$ sub-shell closure, a stronger configurational isospin polarization of the one-phonon states than in $^{94}$Mo is expected~\cite{Werner02,Holt01}. 
Recent work showed that the collectivity of the low-lying symmetric and mixed-symmetric quadrupole excitation in vibrational nuclei originates from the coupling of the giant quadrupole resonance to the dominant valence-space configurations~\cite{Walz01}. 
Experimental evidence in $^{92}$Zr and $^{94}$Mo stems from the observation of a difference of the respective matter-transition radii (deduced from proton scattering) while charge-transition radii (deduced from electron scattering) were found to  be about equal.
The difference results from a sign change of the dominant valence neutron amplitude in MSS with respect to the FSS.
 
The present work provides an in-depth study of the electron scattering results on $^{92}$Zr.
In particular, we discuss a new method for a  model-independent determination of the ratio of the $E2$ transition strengths of fully symmetric and mixed-symmetric one-phonon excitations in heavy vibrational nuclei, which at the same time provides an estimate of the sensitivity to the transition-radius difference between these two states. 
The results are furthermore interpreted in the framework of the QPM (for an introduction to the model see \cite{Soloviev01}).

\section{Experiment}

The experiment has been carried out at the Darmstadt superconducting electron linear accelerator S-DALINAC. 
The LINTOTT spectrometer was used with a focal-plane detector system based on four single-sided silicon detectors, each providing 96 strips with thickness of 500 $\mu$m and a pitch of 650 $\mu$m \cite{Lenhardt01}.
Electrons with an incident beam energy $E_0 = 63$~MeV and beam currents ranging from 0.5 to 1 $\mu$A impinged on a $^{92}$Zr target with an isotopic enrichment of 94.57\% and an areal thickness of 9.75 mg/cm$^2$. 
Data were taken at five different scattering angles $\Theta = 69^{\circ}$, $81^{\circ}$, $93^{\circ}$, $117^{\circ}$, and $165^{\circ}$ covering roughly the first maximum of an $E2$ form factor. 

\begin{figure}[tbh!]
\begin{center}
\includegraphics[width=8cm]{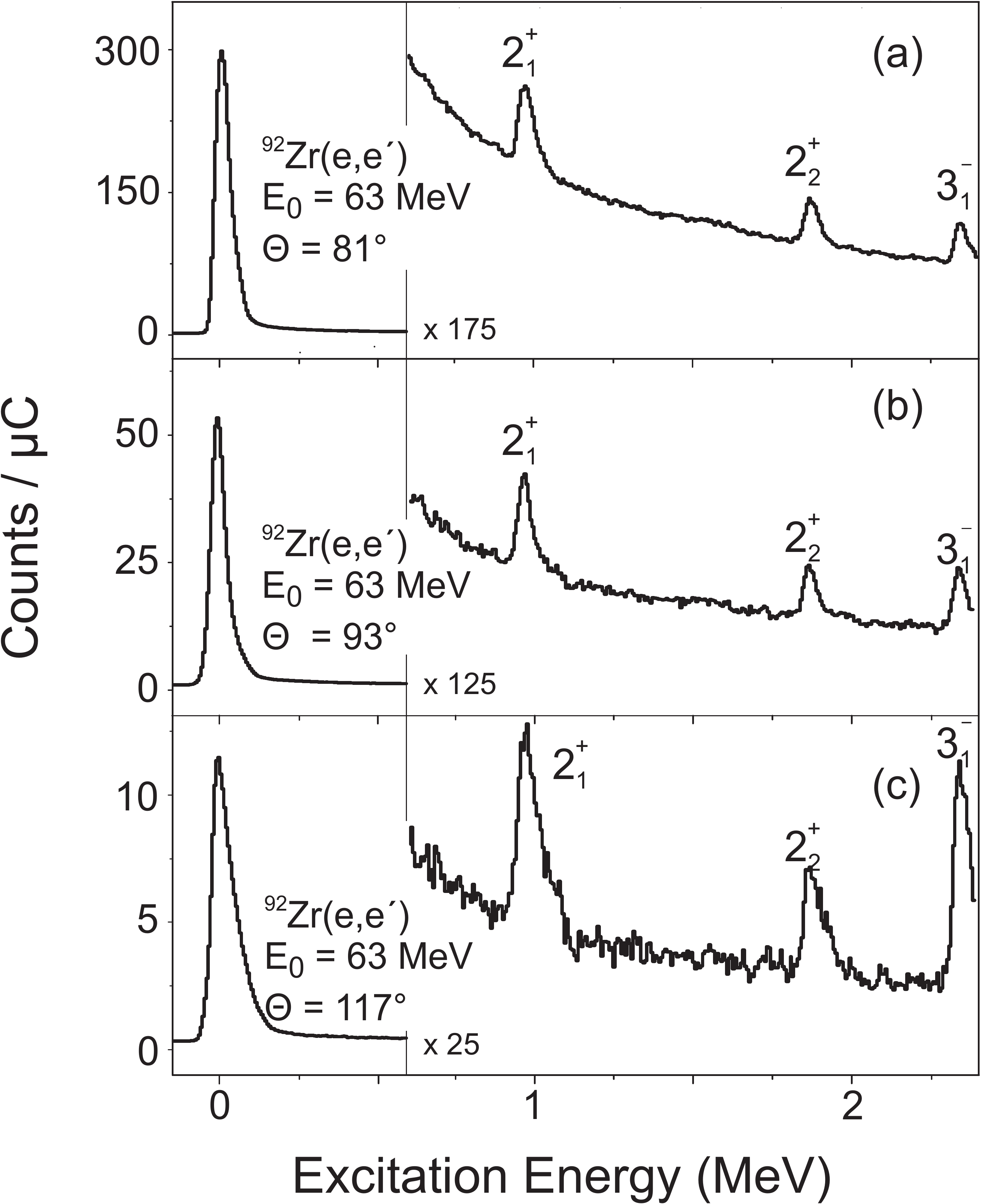}
\caption{Electron scattering spectra of the $^{92}$Zr($e,e'$) reaction at incident electron energy $E_{0} = 63$ MeV and electron scattering angles $\Theta = 81^{\circ}$, $93^{\circ}$ and $117^{\circ}$.}
\label{ZR92_2Spe}
\end{center}
\end{figure}

Examples of electron-scattering spectra are shown in Fig.~\ref{ZR92_2Spe}.
The energy resolution was about 55 keV (full width at half maximum, FWHM).
The observed peaks correspond to the elastic line, the collective one-phonon $2^{+}_{1}$ ($E_{x}=0.934$ MeV) and $3^{-}_{1}$ ($E_{x}=2.339$ MeV) states, and the one-phonon MSS ($2^{+}_{2}$, $E_{\rm x}=1.847$ MeV).
Peak areas $A$ of the transitions were obtained from a spectrum decomposition using the line shape described in Ref.~\cite{Hofmann01}.  
Absolute differential cross sections were determined from normalization to the elastic scattering peak.
Theoretical elastic scattering cross sections were obtained from calculations with the code PHASHI \cite{PHASHI} using charge density distributions from Ref.~\cite{DeJager01}.
The resulting inelastic cross sections with statistical uncertainties are given in Tab.~\ref{tab:ff} normalized to the Mott cross section.
The overall systematic uncertainties of the normalization due to the model description of the charge density and experimental kinematic parameters (electron energy, scattering angle) were estimated to 5\%, which were added in quadrature. 
\begin{ruledtabular}
\begin{table}[t]
\caption{Cross sections of electroexcitation of the $2^+_{1,2}$ and $3^-_1$ states in $^{92}$Zr normalized to the Mott cross section in units $10^{-4}$ and the ratio $R_F$ of the $2^+_{1,2}$ kinematical functions defined in Eq.~(\ref{eqn:ratioDR}).
Only statistical errors are given.}
\label{tab:ff}
\begin{tabular}{c c c c c}
$q$ (fm$^{-1}$) & $2^{+}_{1}$ & $2^{+}_{2}$ & $3^-_1$ & $R_F$ \\
\hline
0.36 &3.98(11)&2.02(7) & 1.95(30) & 1.0148\\
0.41 &5.19(4)&2.67(3) & 3.30(13) & 1.0146 \\
0.46 &5.39(10)&2.95(8) & 4.32(10) & 1.0146 \\
0.55 &7.94(21)&4.23(17) & 10.5(5) & 1.0143 \\
0.64 &5.2(5)&4.4(4) & -       & 1.0143 \\
\end{tabular}
\end{table}
\end{ruledtabular}
\begin{figure}[h]
\begin{center}
\includegraphics[width=8cm]{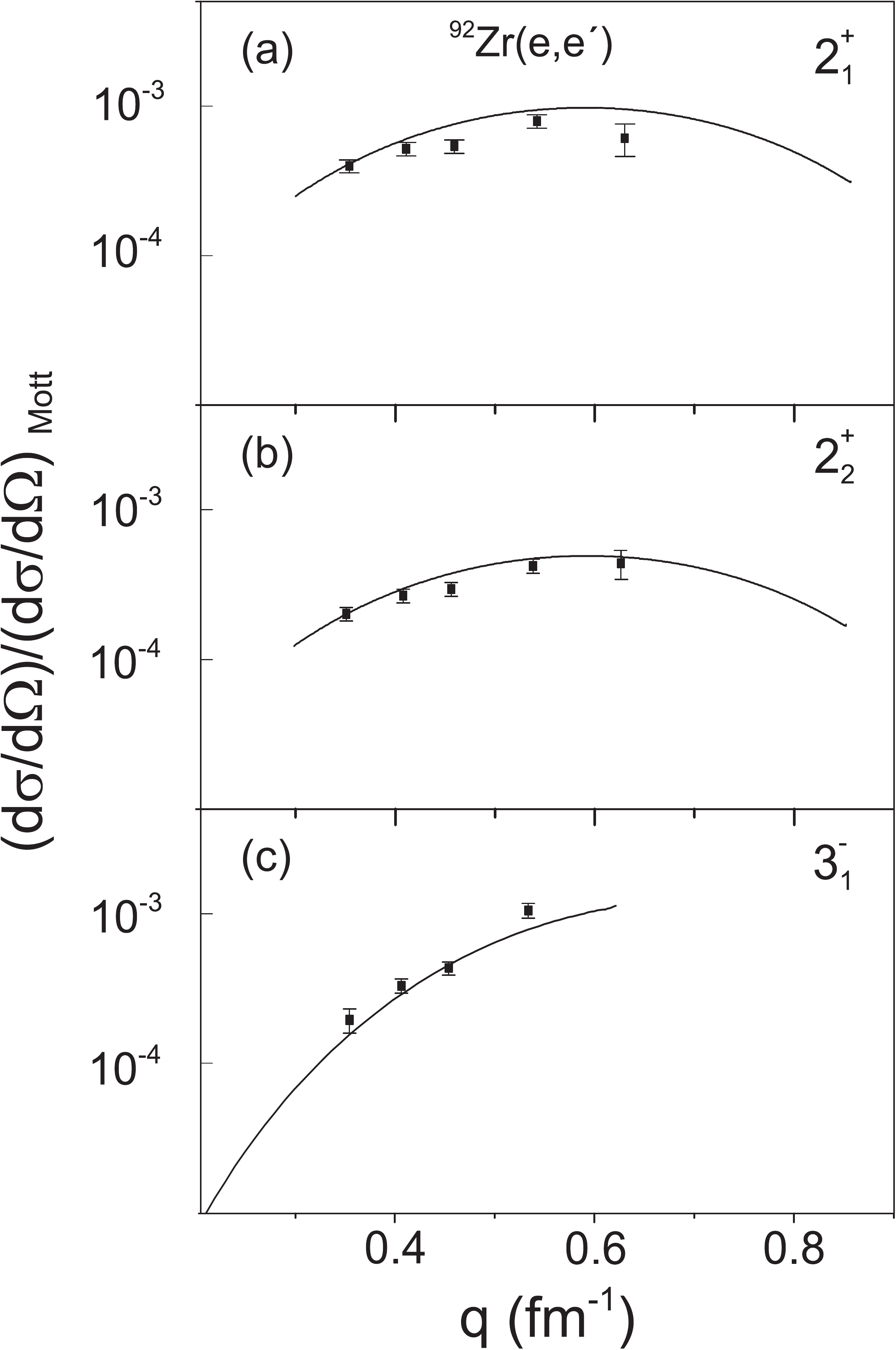}
\caption{Momentum-transfer dependence of the excitation of the one-phonon fully symmetric $2^{+}_{1}$ state (top), mixed-symmetric $2^{+}_2$  (middle) and $3^{-}_{1}$ state (bottom) of $^{92}$Zr from electron scattering. The data (full squares) are compared to the QPM calculations (solid lines).}
\label{2p12p23m1}
\end{center}
\end{figure}
%


%

\section{DWBA Analysis}

Figure \ref{2p12p23m1} presents the results of Tab.~\ref{tab:ff}  in comparison with QPM calculations as a function of momentum transfer
 \begin{equation}
\label{eqn:q}
q = \frac{1}{\hbar c}\sqrt{2E_{\rm 0}\left(E_{\rm 0}-E_{\rm x}\right)\left(1-cos~\theta\right)+E^{\rm 2}_{\rm x}}.
\end{equation} 
In the present application (details are given in Ref.~\cite{Walz01}), excited states in $^{92}$Zr are described by wave functions including one-, two-, and three-phonon configurations. 
Note that the results are very similar to an earlier QPM study of the $2^+$ MSS  in $^{92}$Zr \cite{LoIudice01}. 
Theoretical ($e,e'$) cross sections have been calculated within the Distorted Wave Born Approximation (DWBA) to account for Coulomb distortion effects.
They provide a satisfactory description of  the $q$ dependence.
In order to extract the reduced transition probabilities,  the calculations have been scaled to the data and extrapolated to the photon point, $q \equiv k = E_{x}/\hbar c$.
The results are given in Tab.~\ref{tab:b} labelled "DWBA".
The quoted errors are those of the least-square fit to the data. 
Possible systematic uncertainties due to the extrapolation to the photon point are expected to be negligible.
The absolute $B(E2;2^{+}_{1}\rightarrow0^{+}_{1})$ and $B(E2;2^{+}_{2}\rightarrow0^{+}_{1})$ strengths agree well within error bars with a previous experiment \cite{Fransen02}. 
This is also true for the $B(E3;3^{-}_{1}\rightarrow0^{+}_{1})$ transition strength  but the present value is significantly more precise. 
Previous measurements based on low-energy proton scattering show a large spread of results (14.7, 16.2, 18.9, 21.3, 23.6 W.u.) \cite{Kibedi01}, most likely due to the model dependence of the extraction of an electromagnetic transition matrix element from hadronic scattering data.   

\begin{ruledtabular}
\begin{table}[tbh!]
\caption{Reduced $B(E \lambda$) transition strengths of low-energy collective transitions in $^{92}$Zr deduced from the present $(e,e')$ data in comparison with literature values from Ref.~\cite{Fransen02} for $B(E2$) and Ref.~\cite{Kibedi01} for $B(E3$). The strengths are given in Weisskopf units (W.u.).}
\label{tab:b}
\begin{tabular}{cccc}
 & \multicolumn{2}{c}{Present work}  & Literature \\
 & DWBA & PWBA  & \\
\hline
$B(E2;2^{+}_{1}\rightarrow0^{+}_{1}$) & 6.18(23) & & 6.4(5) \\
$B(E2;2^{+}_{2}\rightarrow0^{+}_{1}$) & 3.31(10) & 3.32(27) & 3.4(4) \\
$B(E3;3^{-}_{1}\rightarrow0^{+}_{1}$) & 18.4(11) & & 19(6)  \\
\end{tabular}
\end{table}
\end{ruledtabular}

\section{PWBA Analysis}

The transition strengths derived from the $(e,e')$ data depend of course on the applied nuclear structure model. 
In light nuclei it has been shown that transition strengths can be extracted in a nearly model-independent plane-wave Born approximation (PWBA) analysis \cite{Theissen01}. 
It assumes that Coulomb distortions of the electron wave function can be approximated by an overall correction factor determined from the g.s.\  charge distribution of the nucleus.  
For kinematics where transverse contributions can be neglected, the differential cross sections for electric transitions are then given by 
\begin{widetext}
\begin{equation}
\label{eqn:PWBA1}
\left(\frac{d\sigma}{d\Omega} \right)_{E\lambda} =f_c\left(\frac{d\sigma}{d\Omega} \right)_{E\lambda, PWBA} =f_c
\frac{\alpha^{2}a_{\lambda} q^{2\lambda}}{k_0^{2}R} 
\frac{\lambda}{\lambda+1} V_{L}(\theta) B(C\lambda, q), \\
\end{equation}
\end{widetext}
%
where
$a_{\lambda}=\pi \lambda^{-1} \left( \lambda + 1 \right) \left[ \left( 2 \lambda + 1 \right) !! \right]^{-2}$, $k_{0} = E_{0}/\hbar c$, and
$R = 1+\hbar c (k_0/M c^{2}) (1-\cos \theta)$. 
The symbol $\alpha$ denotes the fine structure constant, $\lambda$ is the transition multipolarity, and $V_{L}(\theta)$ is a kinematic function given e.g.\ in Ref.~\cite{Theissen01}. 
The quantities B$(C\lambda)$ and the reduced transition probabilities B$(E\lambda)$ from real-photon experiments can be related by Siegert's theorem \cite{Siegert01} 
$$B(C \lambda, q) = q^2/k^2 B(E \lambda, k).$$ 
The Coulomb correction factor 
$$f_{c} (q, E_0, E_{\rm x}) = \left[ \frac{ \left( d\sigma / d\Omega \right)_{\rm DWBA}} {\left(  d\sigma /d\Omega \right)_{\rm PWBA}} \right]_{\rm theo}$$  
is determined from the ratio of DWBA and PWBA calculations employing the QPM transition densities. 

The reduced transition probabilities can thus be related to the experimental differential cross section by 
\begin{eqnarray}
  \label{eqn:PWBA}
      B(C\lambda, q_{\rm x}) & = & \frac{k^{2}_{0}R}{\alpha^{2}a_{\lambda}q^{2\lambda}_{\rm x}}
\left[V_{L}(\theta)\frac{\lambda}{\lambda+1} f_{c}\left(q_{\rm x}, E_{0},E_{x}\right)\right]^{-1} \left(\frac{d\sigma}{d\Omega}\right)_{E\lambda}\\ \nonumber
& \equiv &  \left[f_{\rm kin}f_{\rm c}\left(q_{\rm x}, E_{0},E_{x}\right)\right]^{-1} \left(\frac{d\sigma}{d\Omega}\right)_{E\lambda}.
\end{eqnarray}
%

For low momentum transfers, $B(C\lambda,q)$ can be expanded in a power series of the momentum transfer
\begin{equation}
\label{eqn:expansion}
\sqrt{B (C\lambda,q)} = \sqrt{B \left( C\lambda,0 \right)} \left(1-\frac{q^2 R_{\rm tr}^2}{2\left(2\lambda +3\right)}   + \frac{q^4 R_{\rm tr}^4}{8\left(2\lambda +3\right)\left(2\lambda +5\right)}  -\cdots \right)
\end{equation}
Here, $R^n_{\rm tr}$ is defined by
\begin{equation}
R^n_{\rm tr}= \frac{\displaystyle \langle r^{\lambda+n}\rangle_{\rm tr}} {\displaystyle \langle r^\lambda \rangle _{\rm tr}} = \frac{\displaystyle \int \rho^\lambda_{\rm tr}r^{\lambda + n} d^{3}r} {\displaystyle \int \rho^\lambda_{\rm tr}r^{\lambda} d^{3}r} 
\end{equation}
with $\rho^\lambda_{\rm tr}(r)$ desribing the transition density of multipolarity $\lambda$.

\begin{figure}[tbh!]
\includegraphics[width=8cm]{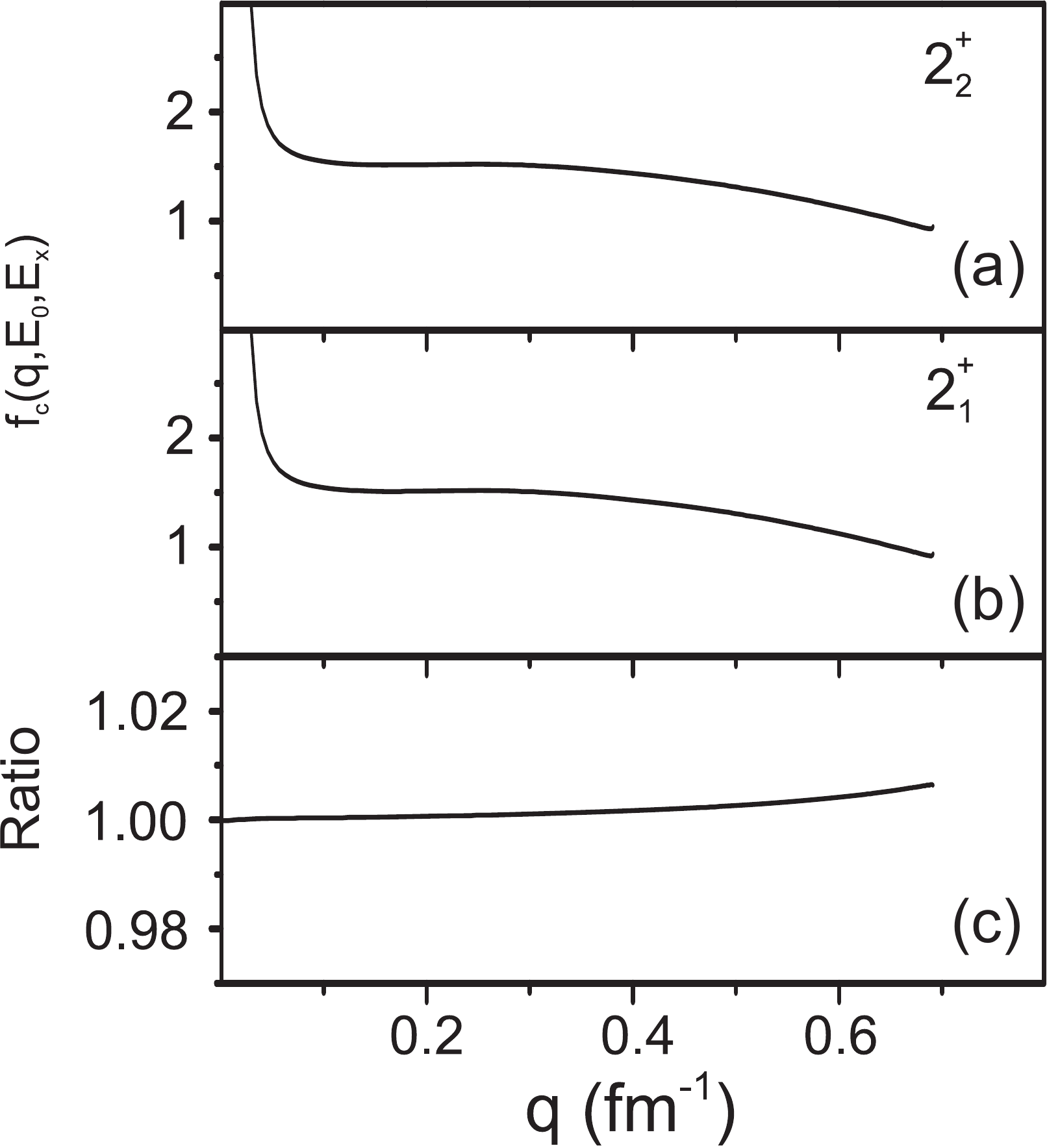}
\caption{Momentum-transfer dependence of the Coulomb corrections for the transition to the $2^+_{1,2}$ states in $^{92}$Zr for an incident electron energy $E_0 = 63$ MeV.}
\label{CoulombRatio}
\end{figure}

In general, the PWBA approximation is not valid for a heavy nucleus like $^{92}$Zr.
However, it may hold for the ratio of  cross sections populating  the $2^+$ FSS and MSS for the following reasons: (i) the kinematics for both transitions are almost identical, and (ii) transition densities of collective transitions of a given multipolarity are similar (see, e.g., Fig.~2 in Ref.~\cite{Walz01} for the cases studied here).   
Figure~\ref{CoulombRatio} shows the Coulomb-correction factors calculated with the QPM for the transitions to $2^{+}_{1}$ (middle) and $2^{+}_{2}$ (top) states in $^{92}$Zr together with their ratio (bottom) as a function of $q$. 
The ratio is unity to better than 1\% over the range of the momentum transfer included in our experiments. 
Consequently, the effects from Coulomb distortion can be neglected in a relative analysis, and the extraction  of the $B(E2)$ ratio can be achieved with improved accuracy, since systematic errors in the determination of absolute cross sections cancel.

Employing Eqs.~(\ref{eqn:PWBA}) and (\ref{eqn:expansion}) and defining the transition radius $R_{\rm tr} =\sqrt{R_{\rm tr}^2}$, the ratio of reduced transition strengths can be approximated by
%
\begin{widetext}
\begin{equation}
\label{eqn:ratioDR}
\sqrt{\frac{B(C2,q_2)}{B(C2,q_1)}} 
 =  R_{\rm F}(q)\sqrt{\frac{A_{2}}{A_{1}}}
\approx  \sqrt{\frac{B(E2,k_2)}{B(E2,k_1)}} \left(
\frac{1- \frac{\displaystyle q_2^2}{\displaystyle 14} \displaystyle \left(R_{{\rm tr},1} + \Delta R \right)^2 + \frac{\displaystyle q_2^4}{\displaystyle 504} \displaystyle \left(R_{{\rm tr},1} + \Delta R \right)^4}{1- \frac{\displaystyle q_1^2}{\displaystyle 14}  \left(R_{{\rm tr},1}\right)^2 + \frac{\displaystyle q_1^4}{\displaystyle 504}  \left(R_{{\rm tr},1}\right)^4} \right),
\end{equation}
 \end{widetext}
where the indices $1,2$ indicate the transitions to the $2^+_1$ and $2^+_2$  state, respectively. 
$R_F$ denotes the ratio of kinematic functions $\sqrt{f_{kin,2}/f_{kin,1}}$, given in Tab.~\ref{tab:ff}, and $\Delta R = R_{{\rm tr},2} - R_{{\rm tr},1}$ the difference of the corresponding charge-transition radii. 
The experimental ratio depends on the square root of the ratio of the peak areas $\sqrt{A_2/A_1}$ only. 

For the approximation on the r.h.s.\ of Eq.~(\ref{eqn:ratioDR}) use is made of Siegert's theorem and of the Tassie model \cite{Tassie01} which provides a good description of the surface behaviour of transition densities for collective states. 
We have checked that the approximation $R^4_{tr} = (R_{tr})^4$ employed in Eq.~(6) within this model, yields very accurate results, and thus it is used in our analysis below. 
This approximation may be questionable for the results at the highest $q$ in Tab.~\ref{tab:ff} which also have poor statistics. 
The data point is thus omitted in the further analysis.

\begin{figure}[tbh!]
\begin{center}
\includegraphics[width=8cm]{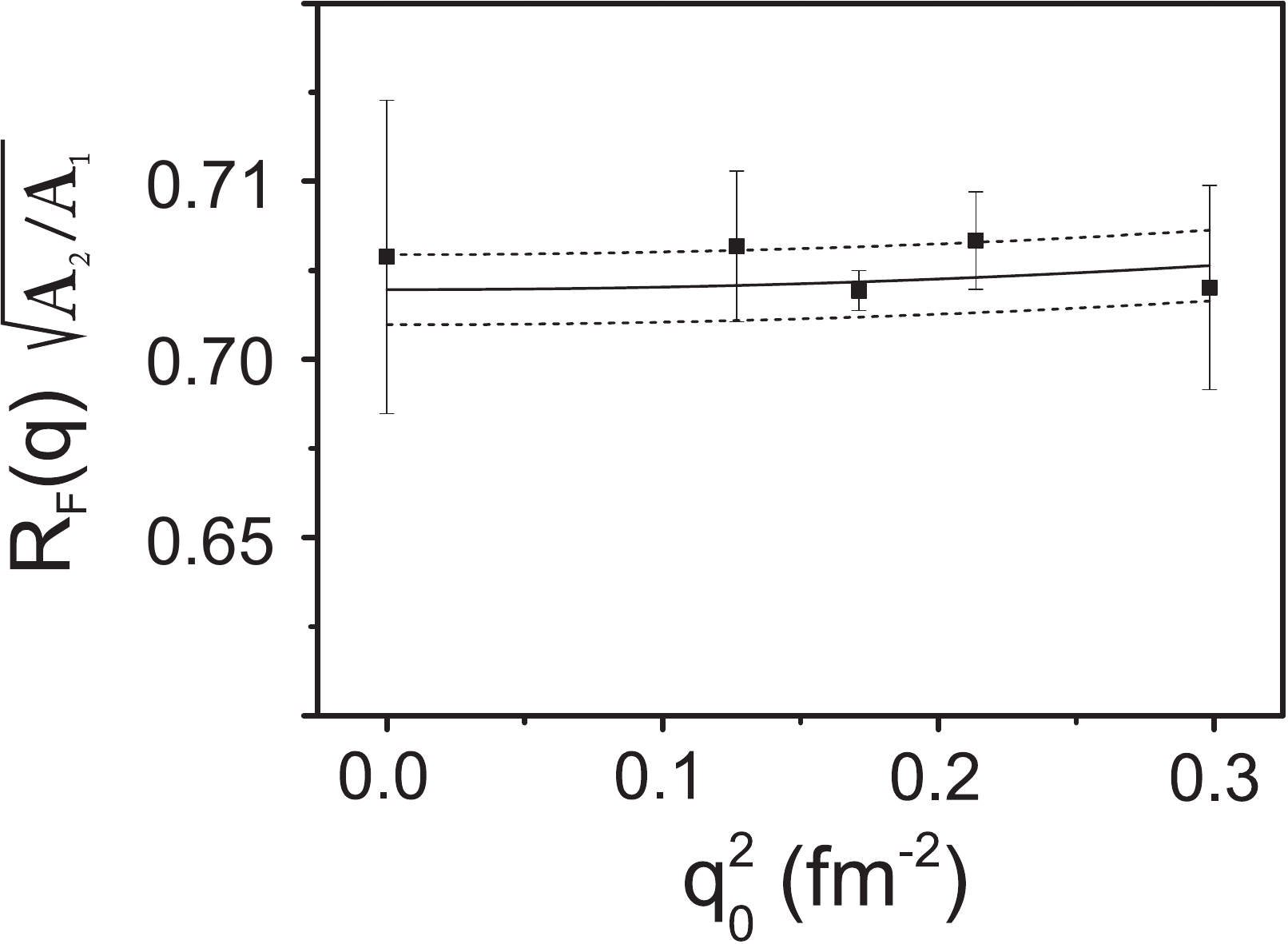}
\caption{Ratio of the reduced transition probabilities of the $2^+$ MSS and FSS (solid squares) of $^{92}$Zr as a function of the squared elasic momentum transfer $q_{0}$. 
An additonal data point (full circle) at $q^2_0 = k^2$ stems from the ratio of $B(E2)$ strengths obtained from $\gamma$-decay lifetime measurements \cite{Fransen02}. 
The solid line is a fit of Eq.~(\ref{eqn:ratioDR}) with $1\sigma$ error bars  given by the dashed lines. }
\label{fitdata}
\end{center}
\end{figure}
Figure~\ref{fitdata} shows a plot of $R_{\rm F}\sqrt{A_2/A_1}$ as a function of the squared elastic momentum transfer.  
A fit of Eq.~(\ref{eqn:ratioDR}) to the data has 3 parameters, viz.\ the ratio of $B(E2)$ strengths, $R_{{\rm tr},1}$ and $\Delta R$.
In a first step , $R_{{\rm tr},1}=5.6$~fm is fixed using the QPM results. 
A $\chi^{2}$-minimization of Eq.(\ref{eqn:ratioDR}) to the data then determines $$\sqrt{\frac{B(E2;2^{+}_{2})}{B(E2;2^{+}_{1})}}=0.720(8).$$ 
With the $B(E2;2^+_1)$ value from Tab.~\ref{tab:b}, we obtain  $B(E2;2^{+}_2)=3.32(27)$~W.u., in agreement with Ref.~\cite{Fransen02} and with the value obtained above from the DWBA analysis.
While the ratio can be determined precisely with an uncertainty of about 1\%, the accuracy of the absolute value is presently limited by  the error of the $B(E2;2^{+}_{1})$ value in the literature. 

The second parameter $\Delta R$ in Eq.~(\ref{eqn:ratioDR}) provides information about the change of the proton transition radii between both $2^{+}$ states. 
This is particular interesting in view of the recent results of Walz {\it et al}.~\cite{Walz01} providing evidence for a significant difference of the neutron transition radii for these two states, while their proton transition radii are expected to be very close based on QPM calculations.
This may serve as a new experimental signature of MSS in vibrational nuclei with a specific shell structure.
An experimental confirmation of this conjecture is of considerable interest.
The fit of Eq.~(\ref{eqn:ratioDR}) leads to $\Delta R = -0.18(65)$ fm, where the uncertainty is dominated by the limited number of data points with small enough error bars at sufficiently low momentum transfers.
One way to improve the fit is the inclusion of the results of Ref.~\cite{Fransen02} providing an additional data point at $q_0 = k$. 
The resulting fit (red curves) with $1\sigma$ error bars (green curves) shown in Fig.~\ref{fitdata} corresponds to $\Delta R = -0.12(51)$ fm consistent with equal proton transition radii to about half a fm.

\begin{figure}
\begin{center}
\includegraphics[width=80mm]{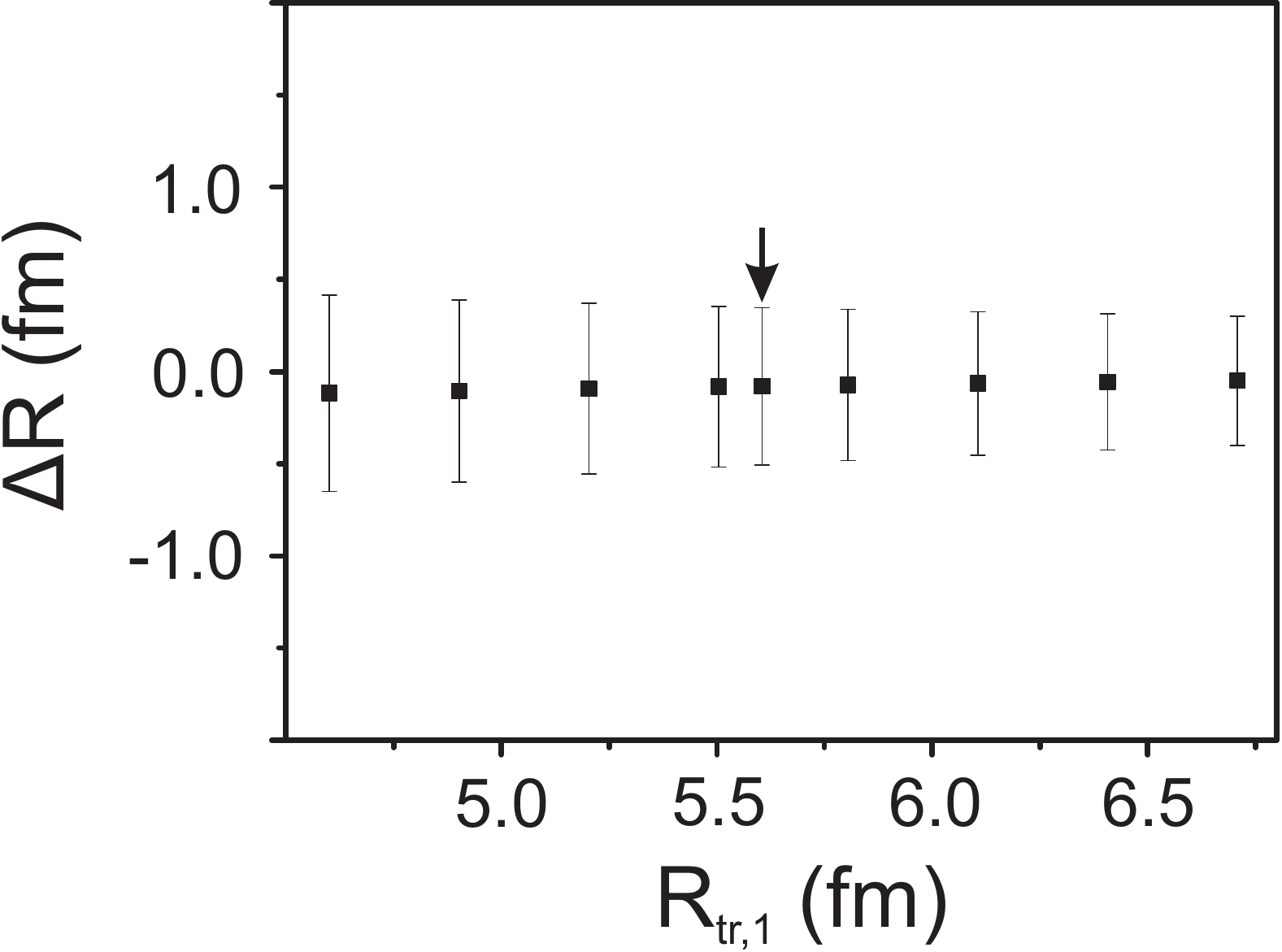}
\caption{Evolution of the charge-transition radii difference between the $2^+_1$ and $2^+_2$ states in $^{92}$Zr obtained from Eq.~(\ref{eqn:ratioDR}) as a function of the transition radius $R_{{\rm tr},1}$. The arrow indicates the prediction of the QPM calculation.}
\label{R9Value}
\end{center}
\end{figure}
Finally, we briefly comment on a possible dependence of the result on a variation of the absolute size of the charge transition radius $R_{{\rm tr},1}$. 
We have repeated the analysis for 9 different values of $R_{{\rm tr},1}$ between 4 and 7 fm, thereby overexhausting the range of possible values expected from model calculations and from the phenomenological finding that the transition radii of collective excitations differ not too much from the radius of the nuclear ground state \cite{Theissen01}. 
As demonstrated in Fig.~\ref{R9Value}, the deduced difference of the charge transition radii is independent of the choice of $R_{{\rm tr},1}$.

 \section{Summary}

To summarize, an investigation of the nature of one-phonon symmetric and mixed-symmetric $2^{+}$ states in $^{92}$Zr has been performed using inelastic electron scattering at low momentum transfers. 
A comparison of the measured form factors with QPM calculations confirms the dominant one-phonon structure of the transitions to the $2^{+}_{1}$ and $2^{+}_{2}$ states.
It is shown that a PWBA analysis of the form factors, which usually fails for heavy nuclei, can nevertheless be applied to extract the ratio of the g.s.\  $B(E2$) transition strengths in a relative analysis.
This is a new promising approach to determine the g.s.\ transition strength of the $2^+$ MSS in vibrational nuclei with a precision limited only by the experimental information about the $B(E2;2^{+}_{1}\rightarrow0^{+}_{1})$ strength.
The PWBA approach furthermore provides information about differences of the proton transition radii of the respective states, containing independent information about the mixed-symmetry character of $2^+$ states and the sign change of leading valence shell components between FSS and MSS \cite{Walz01}.  
For $^{92}$Zr, the proton transition radii agree within about 0.5 fm, consistent with predictions that the sign change arises in this case from the neutron component.

Further analysis of the data indicates that an improved precision for the proton transition radii difference can be achieved by additional data, in particular in momentum transfer ranges presently not covered well (e.g., $q^2 = 0 - 0.1$~fm$^{-2}$).
Elementary to the present approach is an approximate cancellation of Coulomb corrections of the FSS and MSS. 
This may be questioned when moving away from shell closures, where the collectivity of the MSS ground-state decay decreases. 
Systematic investigations are necessary to establish the range of applicability of this new promising method.
Work along these lines (e.g.\ studies of  $^{94,96}$Zr and Mo isotopes) is underway. 

We thank R.~Eichhorn and the S-DALINAC crew for their commitment in delivering electron beams.
This work has been supported by the DFG under grant SFB 634.

\end{document}